\documentclass[10pt]{article}
\usepackage{amsmath}
\usepackage{amssymb}
\usepackage{graphicx}
\usepackage{cite}
\usepackage{color} 
\usepackage[labelfont=bf,labelsep=period,justification=raggedright]{caption}
\makeatletter
\renewcommand{\@biblabel}[1]{\quad#1.}
\makeatother

\date{}

\begin{document}

\begin{flushleft}
{\Large
\textbf{Single-base mismatch profiles for NGS samples}
}
\\
Marco Chierici$^{1}$, 
Giuseppe Jurman$^{1}$,
Marco Roncador$^{1}$,
Cesare Furlanello$^{1,\ast}$
\\
\bf{1} Fondazione Bruno Kessler, Trento, Italy
\\
$\ast$ E-mail: furlan@fbk.eu
\end{flushleft}

\section*{Abstract}
Within the preprocessing pipeline of a Next Generation Sequencing
sample, its set of Single-Base Mismatches is one of the first
outcomes, together with the number of correctly aligned reads. The
union of these two sets provides a $4\times 4$ matrix (called Single
Base Indicator, SBI in what follows) representing a blueprint of the
sample and its preprocessing ingredients such as the sequencer, the
alignment software, the pipeline parameters. In this note we show
that, under the same technological conditions, there is a strong
relation between the SBI and the biological nature of the sample.  To
reach this goal we need to introduce a similarity measure between
SBIs: we also show how two measures commonly used in machine learning
can be of help in this context.

\section*{Introduction}
A first measure of the goodness of alignment of a Next Generation
Sequencing (NGS) sample is given on one side by the number of
correctly aligned reads, and on the other side on the number of
wrongly detected bases, collected in the single-base mismatches
count. All these values are among the basic outcomes of the sample
preprocessing, regardless of the ingredients of the used pipeline.
These information can be grouped together in a $4\times 4$ matrix of
integer values indexed by the four bases A,C,G,T, where the entry in
position X,Y is the number of true bases (in the reference genome) X
interpreted as Y. In what follows, we call this matrix the Single-Base
Indicator, SBI for short, of the studied sample.

Obviously, the SBI is depending not only on the sample, but also on
the adopted pipeline: thus, the sources of variabilities possibily
affecting the final outcome are several. In fact, different alignment
performance have been assessed for different software (Bowtie and BWA
only to mention the two most popular ones), the preprocessing (with
separated or unified lanes) and filtering procedures applied, and the
sequencing platform. Thus the SBI can be seen as (a sort of) signature
for the investigated sample with respect to the employed pipeline
configuration, with the aim of using it to quantitatively formalize
differences; for example, to be used as a baseline reference for the
noise level that may be expected for a given configuration and thus,
in some sense, for the stability of the configuration itself mimicking
what has been carried out in \cite{jurman08algebraic}. To reach this
goal, a similarity measure for SBI is needed: in this paper, we
propose to use concepts from Statistical Machine Learning to deal with
this problem, and namely the theory of classifier performance
comparison.

In fact, the SBI can be seen as the confusion matrix of a
classification problem on the four classes A,G,C,T, where the
classifier is the alignment pipeline and the ground truth is the
reference genome.  Within this framework, several measures are
available in literature for translating a confusion matrix into a
single performance value: see for instance \cite{felkin07comparing,
  sokolova09systematic, ferri09experimental} for a review of the more
classical methods, or \cite{wei10novel} for novel measures or
\cite{landgrebe08efficient, freitas07confusion, freitas07distance} for
totally different approaches. Among others, the Matthews Correlation
Coefficient (MCC, for short) has recently attracted the attention of
many researchers in the field: in particular, it has been designed as
the elective measure in initiatives defining methodologic guidelines
such as MAQC-II \cite{maqc10maqcII} led by the U.S. Food and Drug
Administration (FDA). Originally introduced for binary problems
\cite{matthews75comparison}, its generalization to the multiclass case
was defined in \cite{gorodkin04comparing} and since then used in a
wide range of tasks.  Here the (generalized) MCC is used to summarize
in a unique real value the quality of the alignment of a sample: the
range of the values is $[-1,1]$ and the higher the value the better
the alignment, with 1 representing the ideal situation of no mismatch
occurring.

In addition to the comparison with a common ideal situation, it is
worthwhile to consider the mutual similarities and differences between
SBIs: we perform this by looking at each SBI as a (nominal) histogram,
and use a suitable distance. Among the many available distances
\cite{cha02measuring, cha08taxonomy}, we chose to use the Canberra
distance \cite{lance67mixed}. This is motivated by the fact that,
being defined as the ratio between the (absolute value) of the
difference and the sum of two quantities, the Canberra distance is
intrinsically more susceptible to even slight changes around
zero. This case is indeed the one we are interested in, since when
considering the (frequency) histogram corresponding to a SBI, the
values inherited by the single-base mismatches are very close to zero
because of the preponderance of the occurrences of the exact matches.

As an application of these two similarity measures, we show the
evaluation of the alignment of a NGS dataset where all the components
of the pipeline are fixed, yielding the tissue type as the sole
controlled source of variability. The analysis with both MCC and
Canberra distance leads to the hypothesis that there is a strong
relation between the tissue type and the corresponding SBI structures,
thus representing a first promising step towards the development of a
stability theory for NGS data through the SBI.

\section*{Materials and Methods}
\subsection*{Deep sequencing data}
A NGS data set, denoted here as ``BM1'', was used as
testbed for evaluating the proposed method. The data set, previously
described in \cite{wang08alternative} and available on EMBL-EBI
European Nucleotide Archive (ENA) with accession number SRP000727,
consists of deep transcriptome sequencing (RNA-seq) of 11 human
tissues and 6 cell lines. The biological samples were sequenced on the
Illumina Genome Analyzer (Illumina, Inc., San Diego, CA, USA),
obtaining over 400 million short (32 bp) reads. In case a biological
sample was sequenced over multiple lanes, in the following we refer to
each single lane as a \textit{sample}. The data considered here
include a total of 103 samples (lanes) distributed as in Table
\ref{tab:data_bm1}.

\subsection*{Alignment and postprocessing}
The short reads were mapped to the reference human
genome (UCSC assembly hg18, NCBI Build GRCh36.1), including unordered
sequences and alternate haplotypes, with either BWA 0.5.7
\cite{li09fast} (alignment A1), bowtie 0.12.5
\cite{langmead09ultrafast} (alignment A2), and TopHat 1.1.4
\cite{trapnell09tophat} (alignment A3). The mapping was performed
allowing at most two nucleotide differences (i.e., mismatches or gaps)
over the read length.

For each tissue or cell line and for each alignment method, a
\emph{consensus} sequence of the whole genome (i.e., a genetic
sequence of mapped regions, including variants) was called from read
mappings using SAMtools 0.1.7a \cite{li09sequence} with the model
implemented in MAQ \cite{li08mapping}. Small variants (i.e.,
single-base mismatches) were subsequently called from the consensus
sequences and filtered according to read coverage and mapping quality
constraints. In detail, only variants with a mapping Phred quality
score larger than 30 and covered by at least 3 (to prevent spurious
calls) and at most 100 reads (to avoid calls in regions with very high
depth) were kept, following guidelines commonly adopted in the
literature \cite{wendl08aspects, ng09targeted} and suggested by
Illumina, Inc. for small variant discovery \cite{illumina10seminar}.
Finally, we computed an additional ``virtual sample'' by merging the
alignments of different lanes, when present.

\subsection*{Single Base Indicator}
\begin{figure}[!ht]
\begin{center}
\includegraphics[width=0.9\textwidth]{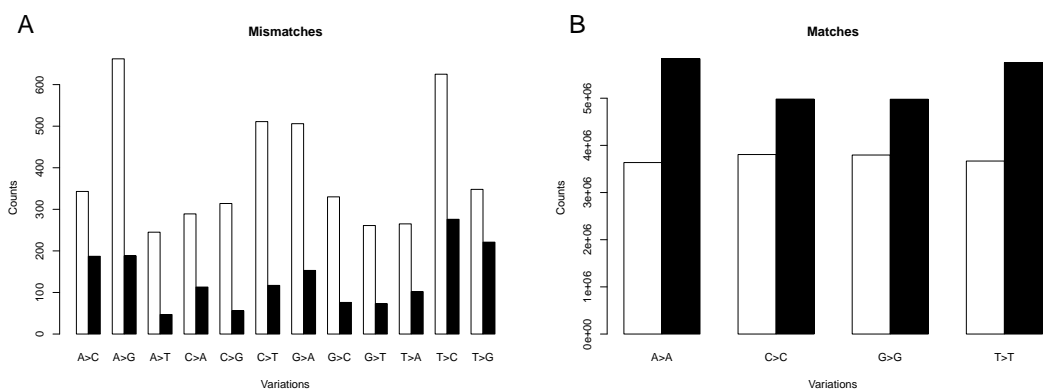}
\end{center}
\caption{
{\bf Alignment A1:} histogram of the SBI for the two BM1 samples SRR015311 (skeletal muscle tissue, in white) and SRR015286 (mixed brain tissue, in black) with different scales for the 12 single-base mismatches categories (A) and the 4 exact matches categories (B).
}
\label{fig:example}
\end{figure}

A typical outcome of any alignment software is a summary of the exact
matches and the occurring single-base mismatches, and these
information may appear under very different shapes. For our purposes,
we are interested in collecting (for each sample, or lane $s$) the 16
integer values $X>Y$ counting the number of times the base $X$ in the
reference genome has been read as $Y$ by the alignemnt pipeline, where
$X,Y\in\mathcal{B}=\{A,C,G,T\}$. The 16 values are then organized in a
square matrix
$\textrm{SBI}(s)\in\mathcal{M}(|\mathcal{B}^2|,\mathbb{N})$, called
Single Base Indicator (SBI), or as a nominal (i.e., with no ordering
among the 16 categories $X>Y$) histogram, possibily normalized over
the total number of counts
$\displaystyle{\textrm{TC}=\sum_{i,j\in\mathcal{B}}
  \textrm{SBI}(s)_{ij}}$. An example of two SBI histograms is shown in
Figure \ref{fig:example}, respectively for the two samples SRR015311
(skeletal muscle tissue) and SRR015286 (mixed brain tissue) in BM1.

\subsection*{Distance from perfect alignment}
Starting from the SBI matrix, define two matrices $X,Y\in\mathcal{M}(\textrm{TC}\times |\mathcal{B}|,\mathbb{F}_2)$ where $X_{sn}=1$ if the base $s$ is predicted to be of class $n$ ($\textrm{pc}(s)=n$) and $X_{sn}=0$ otherwise, and $Y_{sn}=1$ if base $s$ belongs to class $n$ ($\textrm{tc}(s)=n$) and $0$ otherwise. 
Using Kronecker's delta function, the definition becomes:
\begin{displaymath}
X=\left(\delta_{\textrm{pc}(s),n}\right)_{sn}\quad
Y=\left(\delta_{\textrm{tc}(s),n}\right)_{sn}\ .
\end{displaymath}
Then the Matthews Correlation Coefficient MCC can be defined as the ratio:
\begin{displaymath}
\textrm{MCC} = \frac{\textrm{cov}(X,Y)}{\sqrt{{\textrm{cov}(X,X)}\cdot{\textrm{cov}(Y,Y)}}}\ ,
\end{displaymath}
where $\textrm{cov}(\cdot,\cdot)$ is the covariance function. 
In terms of the matrix SBI, the above equation can be written as:
\begin{displaymath}
\frac{\displaystyle{\sum_{k,l,m=1}^N \textrm{SBI}_{kk}\textrm{SBI}_{ml} - \textrm{SBI}_{lk}\textrm{SBI}_{km}}}{
\sqrt{\displaystyle{\sum_{k=1}^N} \left(\displaystyle{\sum_{l=1}^N}\textrm{SBI}_{lk}\right) \left(\displaystyle{\sum_{\substack{f,g=1\\ f\not=k}}^N}\textrm{SBI}_{gf}\right)}
\sqrt{\displaystyle{\sum_{k=1}^N} \left(\displaystyle{\sum_{l=1}^N}\textrm{SBI}_{kl}\right) \left(\displaystyle{\sum_{\substack{f,g=1\\ f\not=k}}^N}\textrm{SBI}_{fg}\right)}
}
\end{displaymath}
MCC lives in the range $[-1,1]$, where $1$ is perfect classification, $-1$ is reached in the alternative extreme misclassification case of a confusion matrix with all zeros but in two symmetric entries $\textrm{SBI}_{\bar i,\bar j}$, $\textrm{SBI}_{\bar j,\bar i}$, and $0$ when the confusion matrix is all zeros but for one single column (all samples have been classified to be of a class $k$), or when all entries are equal $\textrm{SBI}_{ij}=K\in\mathbb{N}$. 
Note that MCC is invariant with respect to multiplication of all SBI's entries by a constant.

\subsection*{Canberra distance between SBIs}
Given the histogram (normalized over the respective TC) of the SBI for
two samples $s,t$ it is possible to define their Canberra distance as
follows:
\begin{displaymath}
\textrm{Ca}(s,t) = \sum_{i,j\in\mathcal{B}} \frac{|\textrm{SBI}(s)_{ij}-\textrm{SBI}(t)_{ij}|}{\textrm{SBI}(s)_{ij}+\textrm{SBI}(t)_{ij}}\ .
\end{displaymath}
Because of the shape of the denominator, variations on the mismatch
categories tend to get higher weight in the sum than the exact
matches, thus highlighting the differences on the wrongly assigned
bases of the two compared samples.

\subsection*{Sammon's mapping}
The projection of SBIs into a two-dimensional space was carried out by
the nonlinear scaling algorithm proposed by
Sammon\cite{sammon69nonlinear}, using $\textrm{Ca}(s,t)$ as the
distance function. The mapping is achieved by minimizing the Sammon
stress function, which in our terms is defined as
\begin{displaymath}
\textrm{E} = \frac{1}{\sum_{s\not=t}\textrm{Ca}^{*}(s,t)} \sum_{s\not=t}\frac{\left(\textrm{Ca}^{*}(s,t) - \textrm{Ca}(s,t)\right)^2}{\textrm{Ca}^{*}(s,t)}
\end{displaymath}
where $\textrm{Ca}^*(s,t)$ is the Canberra distance of samples $s,t$
in the original space and $\textrm{Ca}(s,t)$ is the Canberra distance
in the projected space.

\section*{Results}
\begin{figure}[!ht]
\begin{center}
\includegraphics[width=5in]{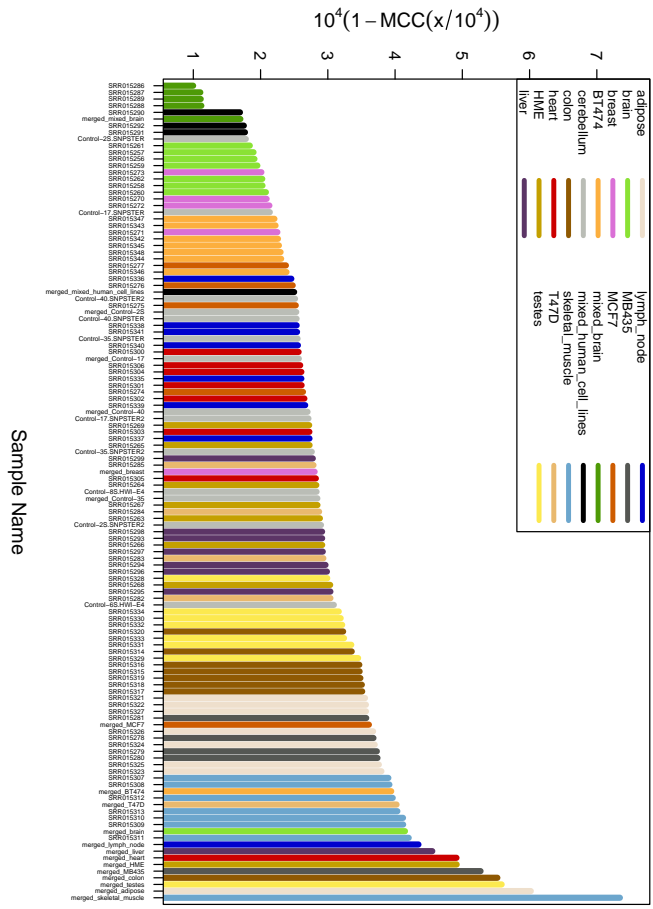}
\end{center}
\caption{
{\bf NGS samples of data set BM1 (alignment A1) ranked according to increasing value of zoomed generalized MCC and colored according to their class:} leftmost samples are closer to the ideal situation of zero mismatches.
}
\label{fig:mcc_bwa}
\end{figure}

\begin{figure}[!ht]
\begin{center}
\includegraphics[width=5in]{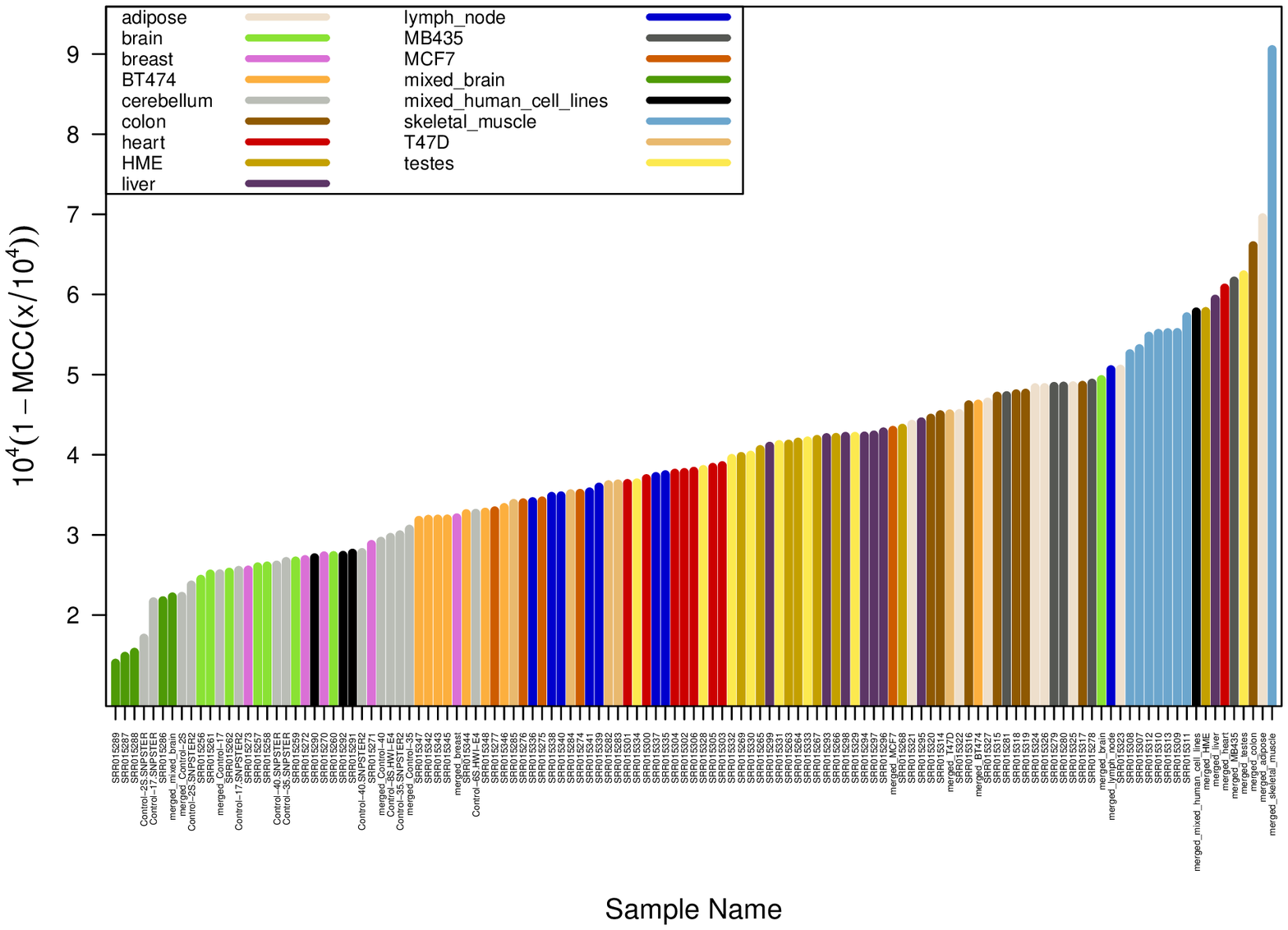}
\end{center}
\caption{
{\bf NGS samples of data set BM1 (alignment A2) ranked according to increasing value of zoomed generalized MCC and colored according to their class:} leftmost samples are closer to the ideal situation of zero mismatches.
}
\label{fig:mcc_bowtie}
\end{figure}

\begin{figure}[!ht]
\begin{center}
\includegraphics[width=5in]{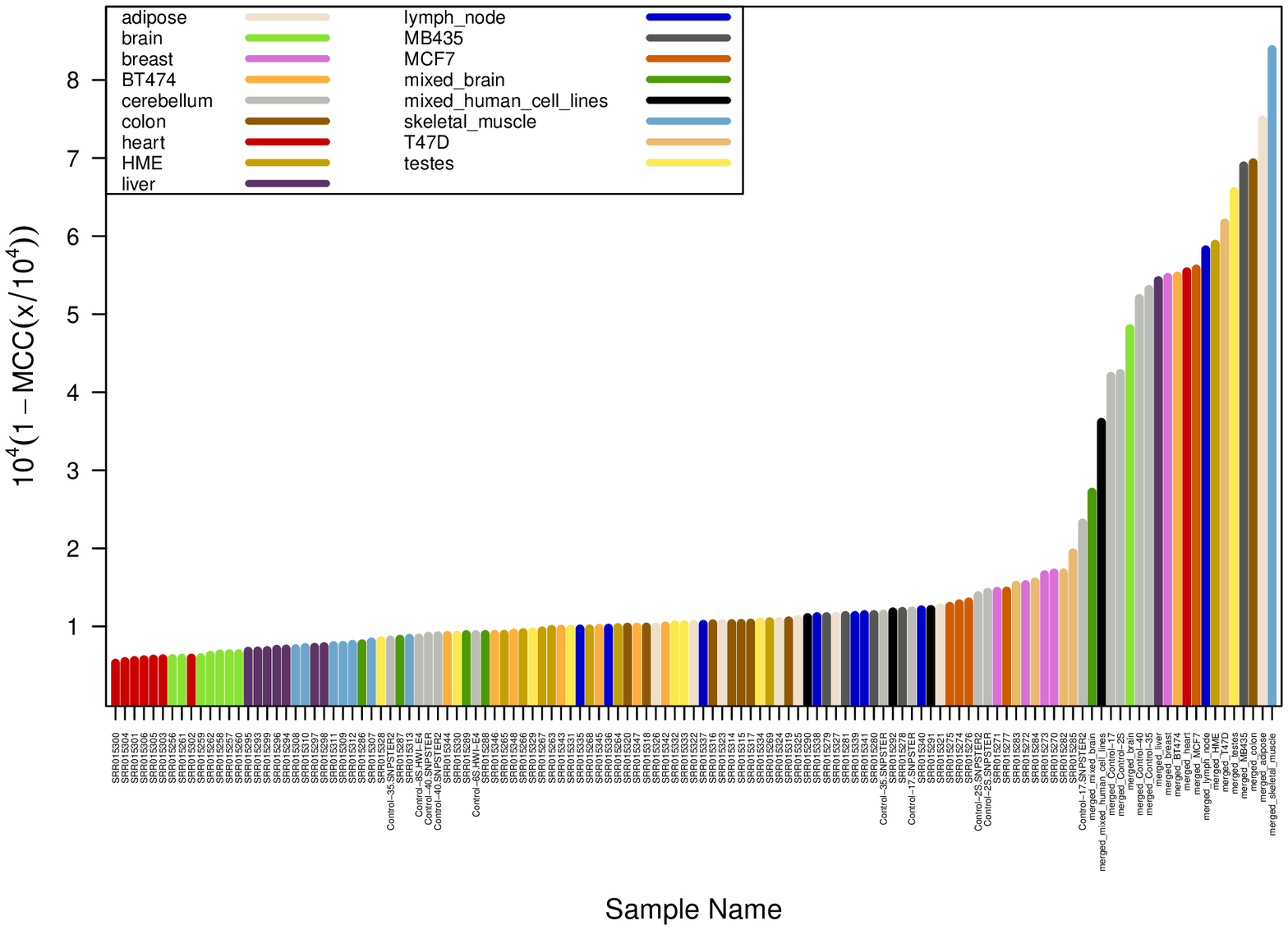}
\end{center}
\caption{
{\bf NGS samples of data set BM1 (alignment A3) ranked according to increasing value of zoomed generalized MCC and colored according to their class:} leftmost samples are closer to the ideal situation of zero mismatches.
}
\label{fig:mcc_tophat}
\end{figure}

The first part of the analysis consists in computing the MCC values
for all the elements of the data set. In order to speed up
computations and to ease data handling, the SBI matrices are loaded in
a portable SQLite database. To better highlight differences, in
Figures \ref{fig:mcc_bwa} (alignment A1), \ref{fig:mcc_bowtie}
(alignment A2) and \ref{fig:mcc_tophat} (alignment A3) we ordered the
123 samples (including 24 virtual samples) of data set BM1 for
decreasing MCC and displayed on the $y$ axes the quantity
$10^4(1-\textrm{MCC})$; thus, the smaller the numbers (leftmost
samples), the closer the SBI to the perfect alignment case. The main
observation here is that samples belonging to the same class tend to
have very similar MCC values, and thus to group together in the plot:
in Tables \ref{tab:pos_bwa}, \ref{tab:pos_bowtie} and
\ref{tab:pos_tophat} we list the rank of all BM1 samples grouped by
class and sorted in decreasing MCC order, for alignments A1, A2 and
A3. The samples of the brain class are the best aligned in average,
with mixed brain (whose tissue is biogically close to brain tissue)
ranking first for A1 and A2. On the other hand, almost all the merged
virtual samples have the worst alignment quality in terms of number of
single-base mismatches.

\begin{figure}[!ht]
\begin{center}
\includegraphics[width=5in]{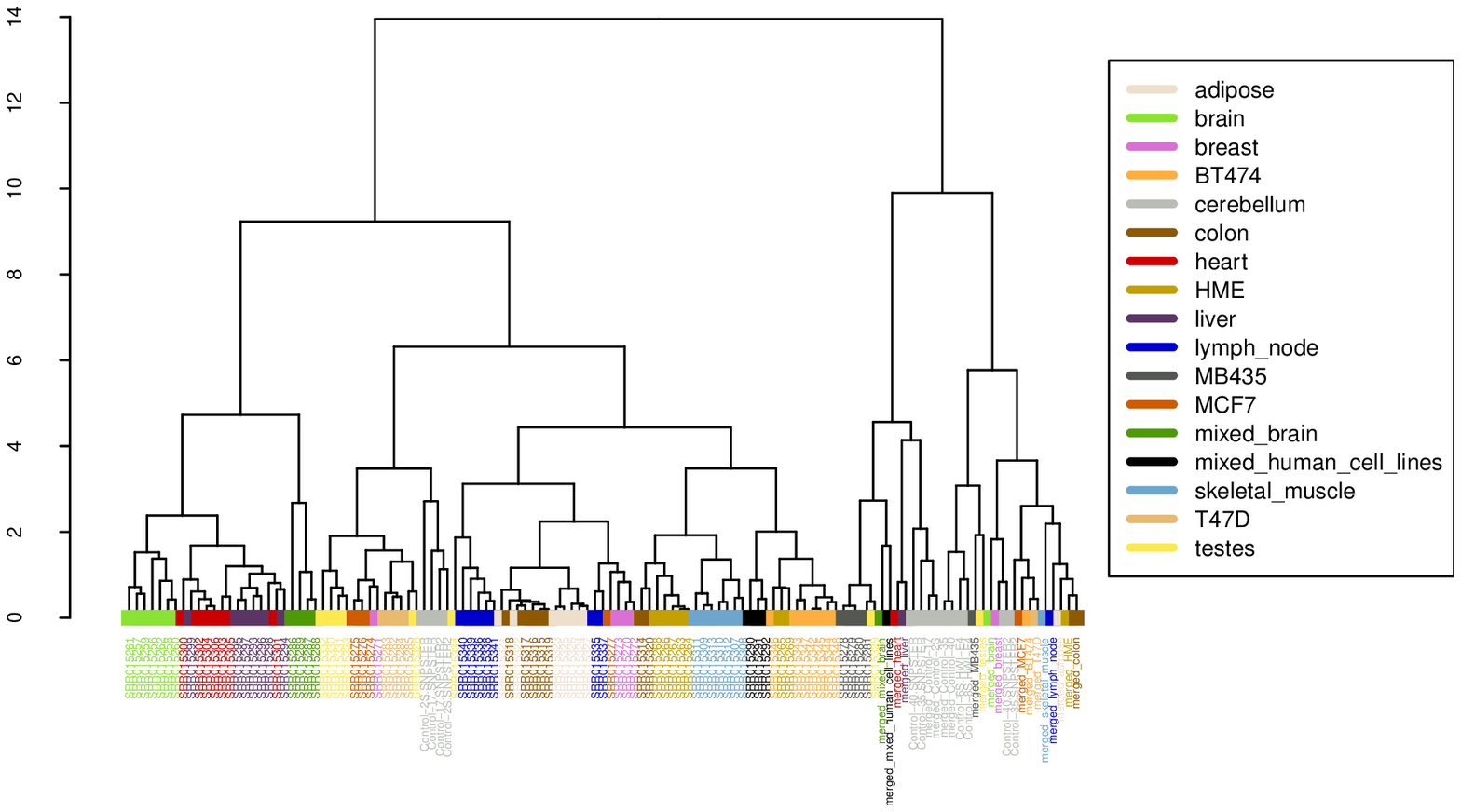}
\end{center}
\caption{
{\bf Data set BM1, alignment A1:} dendrogram of 123 NGS samples (colored according to their class) hierarchically clustered (complete linkage) by Canberra distance.
}
\label{fig:cluster_bwa}
\end{figure}

\begin{figure}[!ht]
\begin{center}
\includegraphics[width=5in]{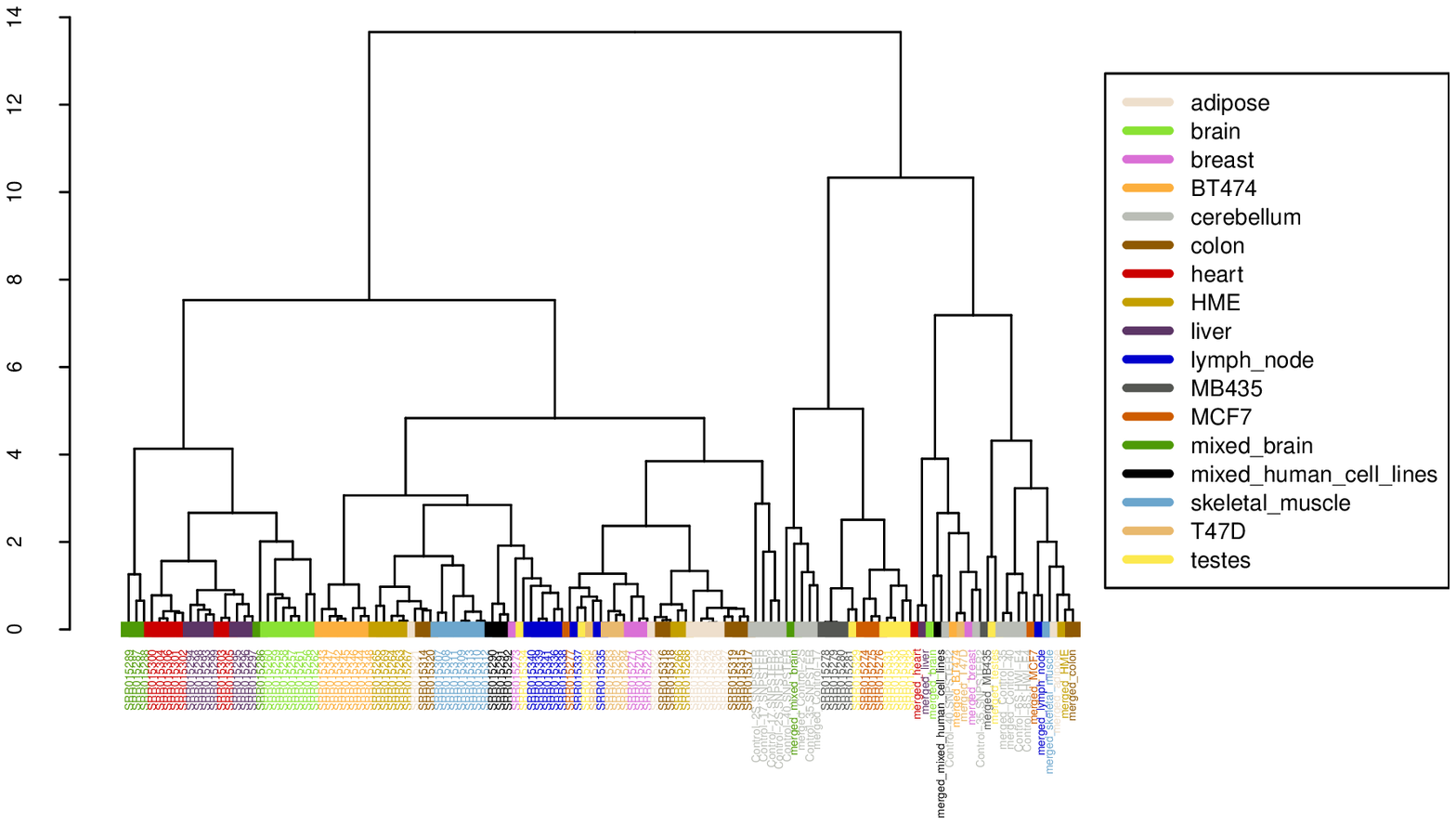}
\end{center}
\caption{
{\bf Data set BM1, alignment A2:} dendrogram of 123 NGS samples (colored according to their class) hierarchically clustered (complete linkage) by Canberra distance.
}
\label{fig:cluster_bowtie}
\end{figure}

\begin{figure}[!ht]
\begin{center}
\includegraphics[width=5in]{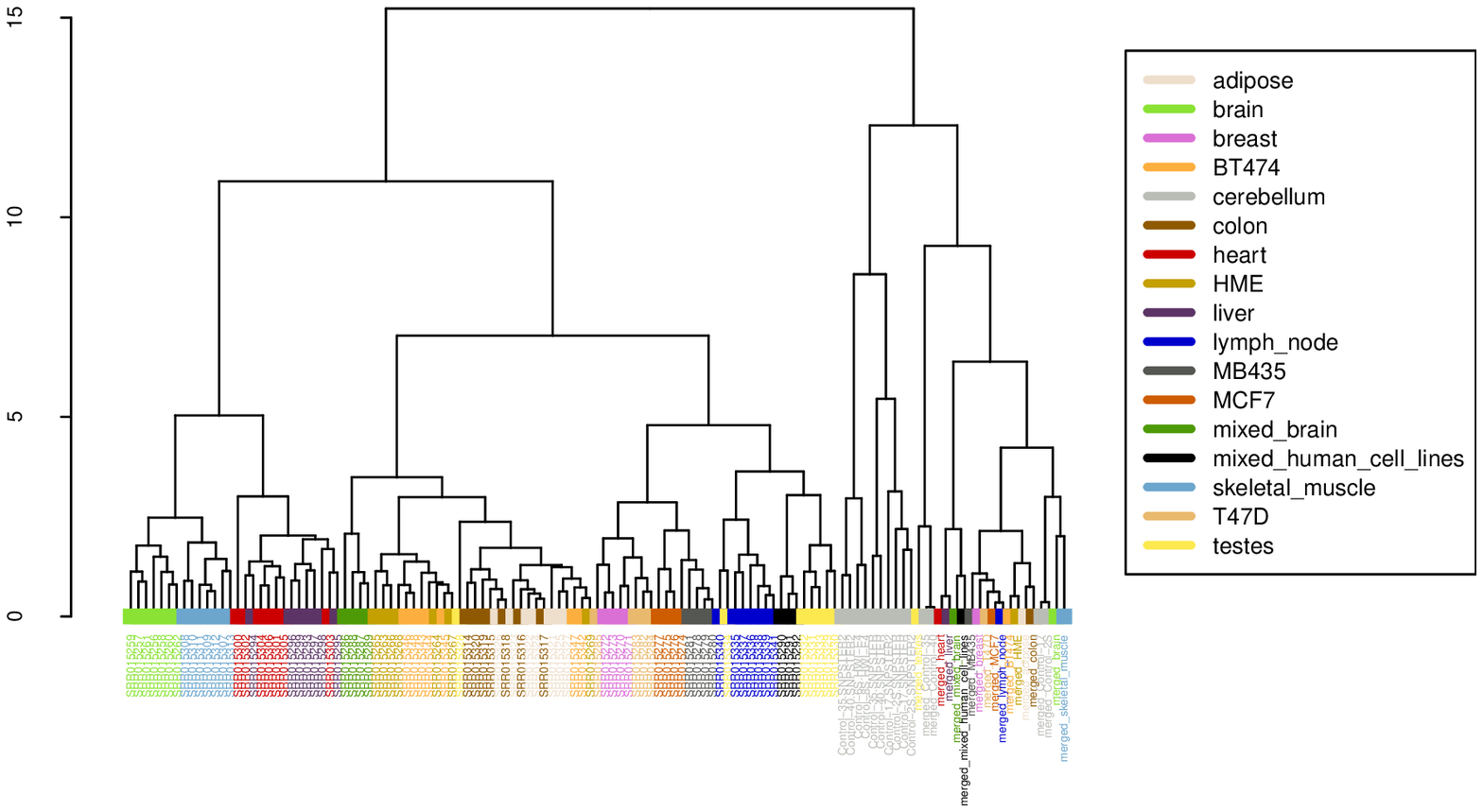}
\end{center}
\caption{
{\bf Data set BM1, alignment A3:} dendrogram of 123 NGS samples (colored according to their class) hierarchically clustered (complete linkage) by Canberra distance.
}
\label{fig:cluster_tophat}
\end{figure}

These claims are furthermore supported by the dendrogram in Figure
\ref{fig:cluster_bwa}, where the mutual 123 Canberra distances among
SBI histograms from alignment A1 are hierarchically clustered together
with complete linkage: Table \ref{tab:cluster_bwa} summarizes the
structure of the clusters resulting by cutting the dendrogram at
height 2. Similar results are reported for alignments A2
(Fig.~\ref{fig:cluster_bowtie}, Tab.~\ref{tab:cluster_bowtie}) and A3
(Fig.~\ref{fig:cluster_tophat}, Tab.~\ref{tab:cluster_tophat}).
Again, for many classes, samples in the same class are consistently
grouped together, with the merged lanes virtual samples forming a
separate entity.

In both analyses, samples of the same class tend to quantitatively
show an intrinsinc similarity. This result seems to support our claim
that the SBI mismatch profile associated to a sample is strongly
dependent on its tissue type, when all other sources of variability
within the preprocessing pipeline are kept stable.

\begin{figure}[!ht]
\begin{center}
\includegraphics[width=5in]{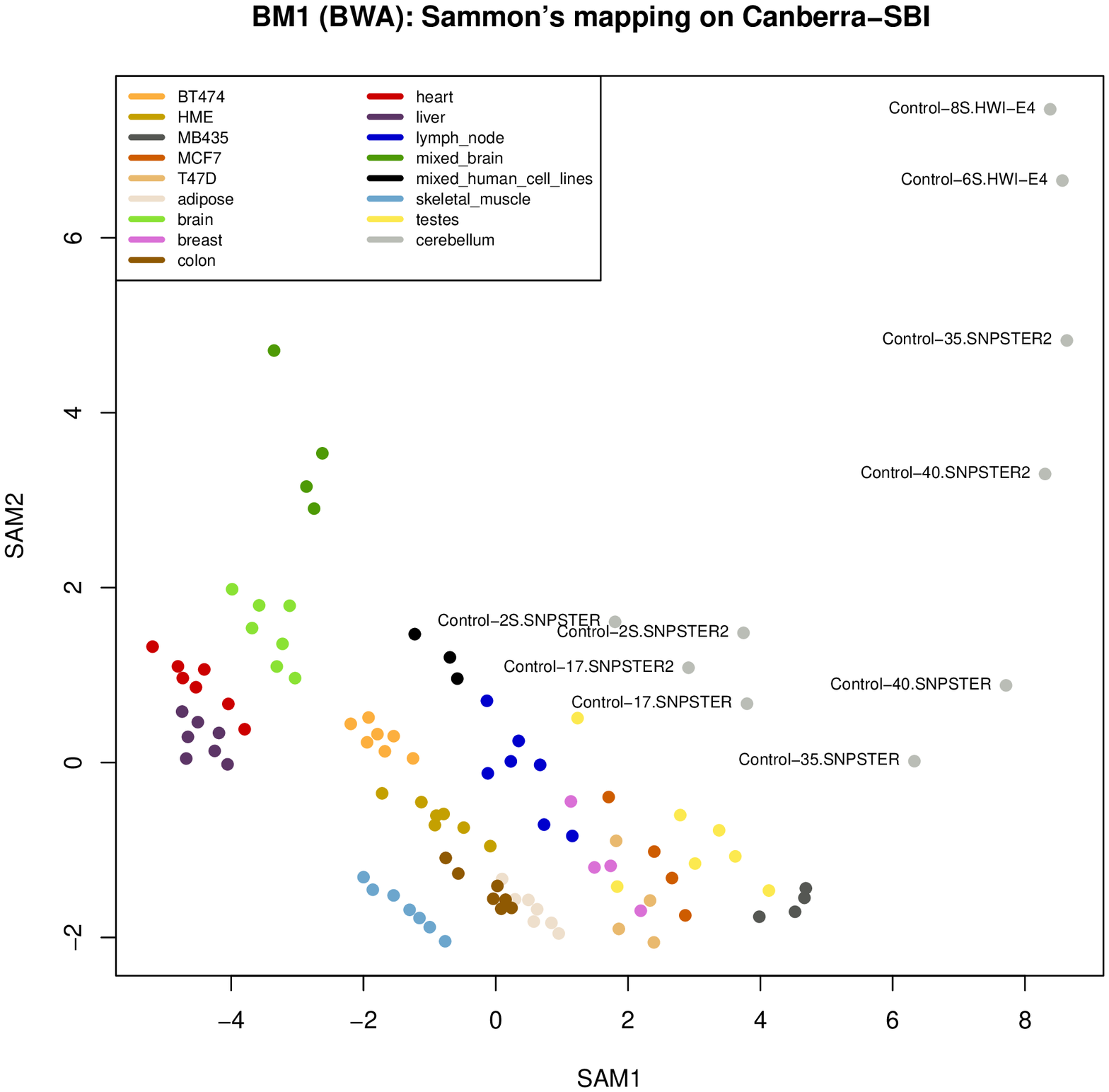}
\end{center}
\caption{
{\bf Sammon map of BM1 samples (alignment A1) using Canberra distance between SBIs:} cerebellum samples are highlighted.
}
\label{fig:bwa_sammon}
\end{figure}

\begin{figure}[!ht]
\begin{center}
\includegraphics[width=5in]{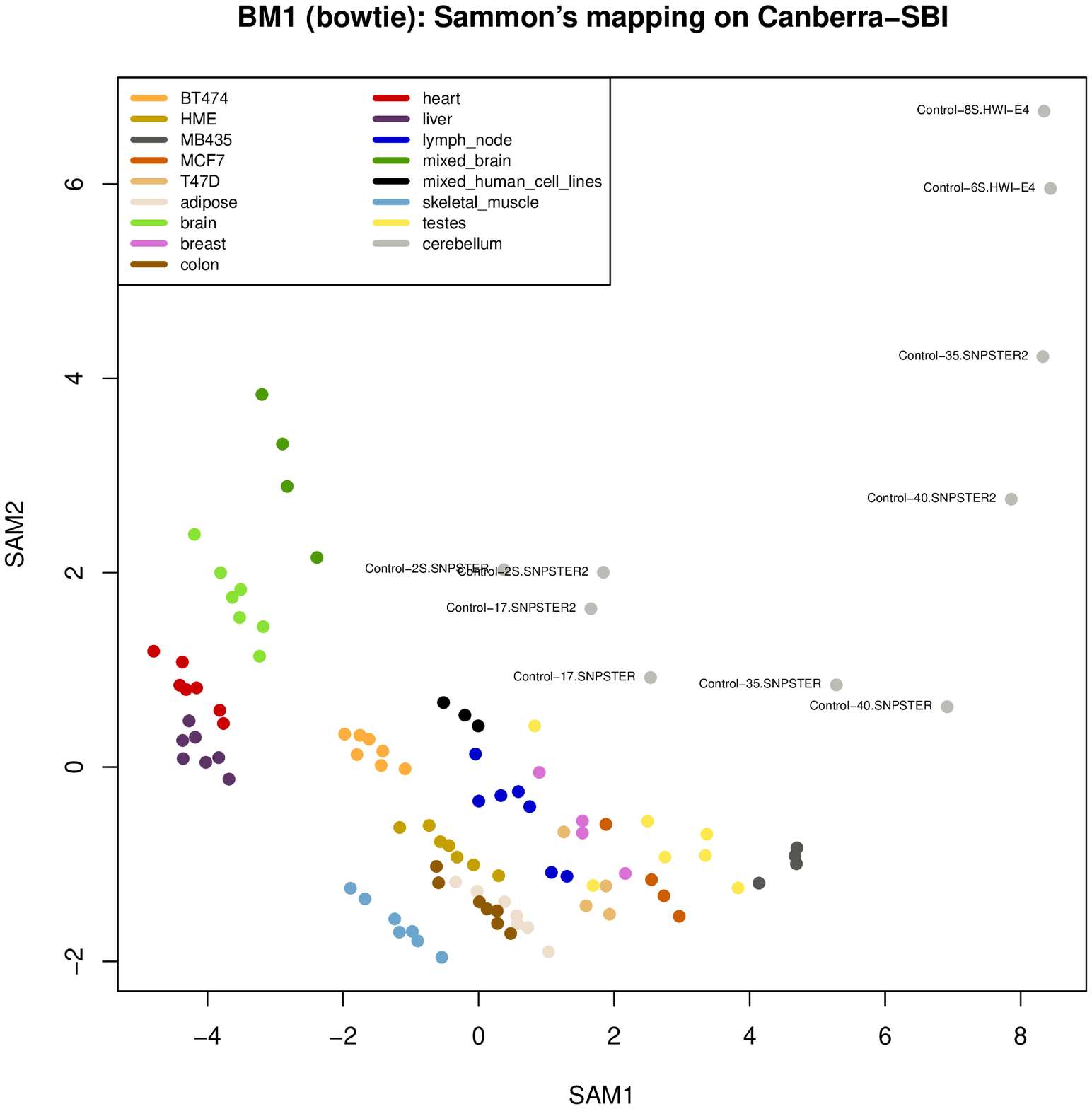}
\end{center}
\caption{
{\bf Sammon map of BM1 samples (alignment A2) using Canberra distance between SBIs:} cerebellum samples are highlighted.
}
\label{fig:bowtie_sammon}
\end{figure}

\begin{figure}[!ht]
\begin{center}
\includegraphics[width=5in]{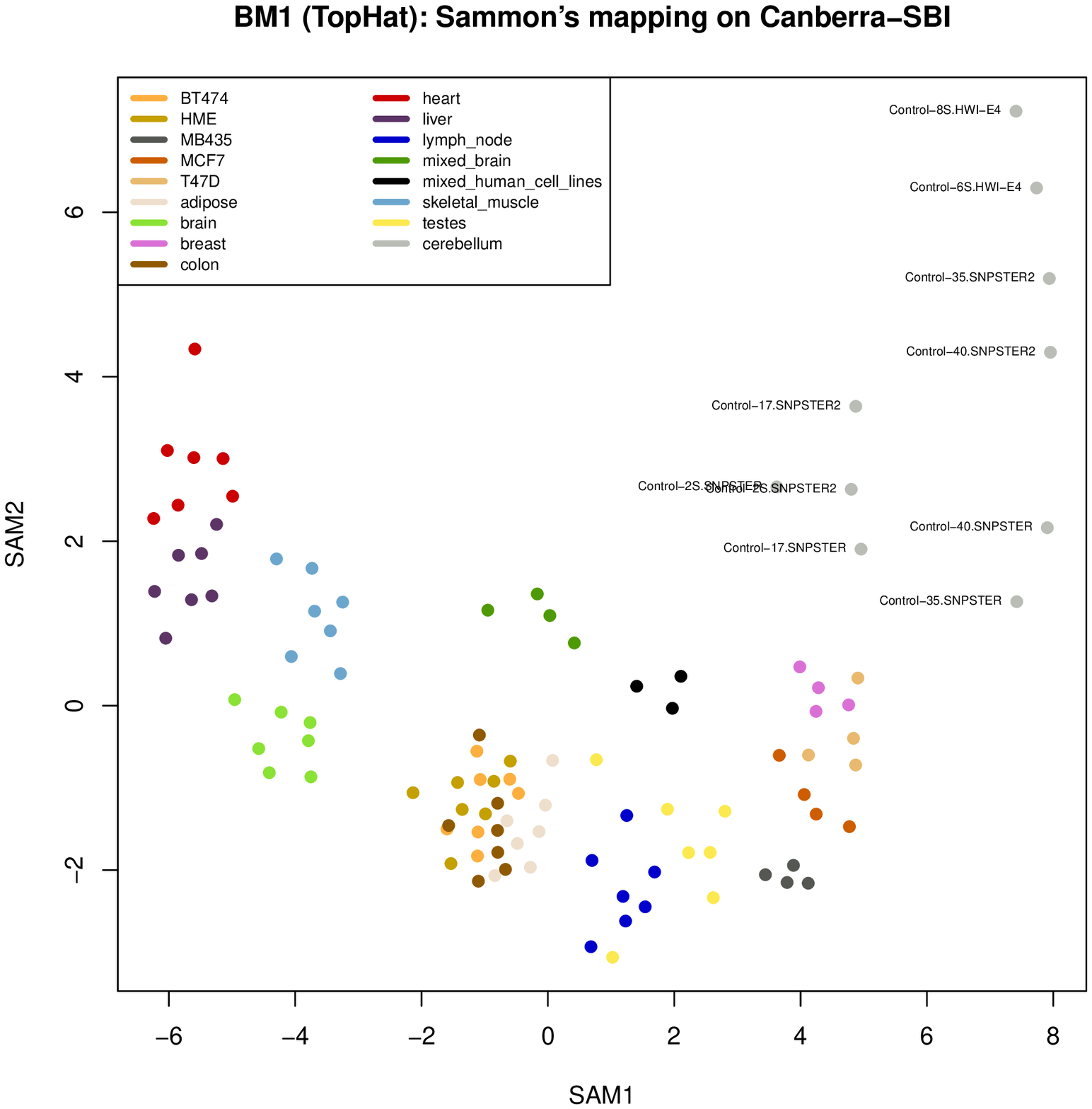}
\end{center}
\caption{
{\bf Sammon map of BM1 samples (alignment A3) using Canberra distance between SBIs:} cerebellum samples are highlighted.
}
\label{fig:tophat_sammon}
\end{figure}

Sammon's mapping of all BM1 samples using Canberra distance between
SBIs (Figures \ref{fig:bwa_sammon}, \ref{fig:bowtie_sammon},
\ref{fig:tophat_sammon}) shows how technical replicates form
well-defined clusters stratified by tissue types, while a separate
trend exists for the biological replicates (cerebellum samples), as
expected. 

\section*{Conclusions}
We propose candidate indicators for profiling RNA-seq experiments
based on generalized MCC and Canberra distance. The goal is to assess
whether variability in RNA-Seq experiments depends on factors that
resemble the analogous of ``batch effects'' for microarrays.

Indicators are built on top of single-base mismatches, which are a
first direct measurement of the goodness of alignment. Results show
that the proposed indicators make it possible to identify
(sub)structures of data determined by underlying experimental
characteristics such as tissue types, paving the way for a
normalization method for NGS that could guarantee the reproducibility
of results, as being nowadays explored by initiatives such as the
Sequencing Quality Control (SEQC) led by U.S. FDA.

The study is currently being extended to consider multiple NGS
platforms, such as Illumina, Helicos, and SOLiD.

\section*{Acknowledgments}
We are grateful to Samantha Riccadonna for her help with the R
statistical environment \cite{r10r}. 

\bibliography{chierici11singleBase}

\begin{thebibliography}{10}
\providecommand{\url}[1]{\texttt{#1}}
\providecommand{\urlprefix}{URL }
\expandafter\ifx\csname urlstyle\endcsname\relax
  \providecommand{\doi}[1]{doi:\discretionary{}{}{}#1}\else
  \providecommand{\doi}{doi:\discretionary{}{}{}\begingroup
  \urlstyle{rm}\Url}\fi
\providecommand{\bibAnnoteFile}[1]{%
  \IfFileExists{#1}{\begin{quotation}\noindent\textsc{Key:} #1\\
  \textsc{Annotation:}\ \input{#1}\end{quotation}}{}}
\providecommand{\bibAnnote}[2]{%
  \begin{quotation}\noindent\textsc{Key:} #1\\
  \textsc{Annotation:}\ #2\end{quotation}}
\providecommand{\eprint}[2][]{\url{#2}}

\bibitem{jurman08algebraic}
Jurman G, Merler S, Barla A, Paoli S, Galea A, et~al. (2008) Algebraic
  stability indicators for ranked lists in molecular profiling.
\newblock Bioinformatics 24: 258--264.
\bibAnnoteFile{jurman08algebraic}

\bibitem{felkin07comparing}
Felkin M (2007) {Comparing Classification Results between N-ary and Binary
  Problems}, Springer-Verlag, volume~43.
\newblock pp. 277--301.
\bibAnnoteFile{felkin07comparing}

\bibitem{sokolova09systematic}
Sokolova M, Lapalme G (2009) A systematic analysis of performance measures for
  classification tasks.
\newblock Information Processing and Management 45: 427--437.
\bibAnnoteFile{sokolova09systematic}

\bibitem{ferri09experimental}
Ferri C, Hern\'{a}ndez-Orallo J, Modroiu R (2009) An experimental comparison of
  performance measures for classification.
\newblock Pattern Recognition Letters 30: 27--38.
\bibAnnoteFile{ferri09experimental}

\bibitem{wei10novel}
Wei JM, Yuan XJ, Hu QH, Wang SQ (2010) A novel measure for evaluating
  classifiers.
\newblock Expert Systems with Applications 37: 3799--3809.
\bibAnnoteFile{wei10novel}

\bibitem{landgrebe08efficient}
Landgrebe TC, Duin RP (2008) {Efficient multiclass ROC approximation by
  decomposition via confusion matrix perturbation analysis}.
\newblock IEEE Transactions Pattern Analysis Machine Intelligence 30: 810--822.
\bibAnnoteFile{landgrebe08efficient}

\bibitem{freitas07confusion}
Freitas COA, De~Carvalho JM, Oliveira JJ Jr, Aires SBK, Sabourin R (2007)
  Confusion matrix disagreement for multiple classifiers.
\newblock In: Rueda L, Mery D, Kittler J, editors, Proceedings of 12th
  Iberoamerican Congress on Pattern Recognition, CIARP 2007, LNCS 4756.
  Springer-Verlag, pp. 387--396.
\bibAnnoteFile{freitas07confusion}

\bibitem{freitas07distance}
Freitas COA, De~Carvalho JM, Oliveira JJ Jr, Aires SBK, Sabourin R (2007)
  {Distance-based Disagreement Classifiers Combination}.
\newblock In: Proceedings of the International Joint Conference on Neural
  Networks, IJCNN 2007. IEEE, pp. 2729--2733.
\bibAnnoteFile{freitas07distance}

\bibitem{maqc10maqcII}
{The MicroArray Quality Control (MAQC) Consortium} (2010) {The MAQC-II Project:
  A comprehensive study of common practices for the development and validation
  of microarray-based predictive models}.
\newblock Nature Biotechnology 28: 827--838.
\bibAnnoteFile{maqc10maqcII}

\bibitem{matthews75comparison}
Matthews BW (1975) {Comparison of the predicted and observed secondary
  structure of T4 phage lysozyme}.
\newblock Biochimica et Biophysica Acta - Protein Structure 405: 442--451.
\bibAnnoteFile{matthews75comparison}

\bibitem{gorodkin04comparing}
Gorodkin J (2004) Comparing two k-category assignments by a k-category
  correlation coefficient.
\newblock Computational Biology and Chemistry 28: 367--374.
\bibAnnoteFile{gorodkin04comparing}

\bibitem{cha02measuring}
Cha SH, Srihari S (2002) {On Measuring the Distance between Histograms}.
\newblock Pattern Recognition 35: 1355--1370.
\bibAnnoteFile{cha02measuring}

\bibitem{cha08taxonomy}
Cha SH (2008) {Taxonomy of Nominal Type Histogram Distance Measures}.
\newblock In: Proc. American Conference on Applied Mathemathics, MATH08. WSEAS,
  pp. 325--330.
\bibAnnoteFile{cha08taxonomy}

\bibitem{lance67mixed}
Lance G, Williams W (1967) {{Mixed-{D}ata {C}lassificatory {P}rograms {I} -
  {A}gglomerative {S}ystems}}.
\newblock Aust Comput J 1: 15--20.
\bibAnnoteFile{lance67mixed}

\bibitem{wang08alternative}
Wang ET, Sandberg R, Luo S, Khrebtukova I, Zhang L, et~al. (2008) {Alternative
  isoform regulation in human tissue transcriptomes}.
\newblock Nature 456: 470--476.
\bibAnnoteFile{wang08alternative}

\bibitem{li09fast}
Li H, Durbin R (2009) {Fast and accurate short read alignment with
  Burrows-Wheeler transform}.
\newblock Bioinformatics 25: 1754--1760.
\bibAnnoteFile{li09fast}

\bibitem{langmead09ultrafast}
Langmead B, Trapnell C, Pop M, Salzberg SL (2009) {Ultrafast and
  memory-efficient alignment of short DNA sequences to the human genome}.
\newblock Genome Biology 10: R25.
\bibAnnoteFile{langmead09ultrafast}

\bibitem{trapnell09tophat}
Trapnell C, Pachter L, Salzberg SL (2009) {TopHat: discovering splice junctions
  with RNA-Seq}.
\newblock Bioinformatics 25: 1105.
\bibAnnoteFile{trapnell09tophat}

\bibitem{li09sequence}
Li H, Handsaker B, Wysoker A, Fennell T, Ruan J, et~al. (2009) {The sequence
  alignment/Map format and SAMtools}.
\newblock Bioinformatics 25: 2078--2079.
\bibAnnoteFile{li09sequence}

\bibitem{li08mapping}
Li H, Ruan J, Durbin R (2008) {Mapping short DNA sequencing reads and calling
  variants using mapping quality scores}.
\newblock Genome research 18: 1851--1858.
\bibAnnoteFile{li08mapping}

\bibitem{wendl08aspects}
Wendl MC, Wilson RK (2008) {Aspects of coverage in medical DNA sequencing}.
\newblock BMC bioinformatics 9: 239.
\bibAnnoteFile{wendl08aspects}

\bibitem{ng09targeted}
Ng SB, Turner EH, Robertson PD, Flygare SD, Bigham AW, et~al. (2009) {Targeted
  capture and massively parallel sequencing of 12 human exomes}.
\newblock Nature 461: 272--276.
\bibAnnoteFile{ng09targeted}

\bibitem{illumina10seminar}
Winn-Deen E (2010).
\newblock A platform approach to translational cancer discovery and diagnostic
  development.
\newblock Oral presentation at Illumina\textit{Dx} Seminar Series.
\bibAnnoteFile{illumina10seminar}

\bibitem{sammon69nonlinear}
Sammon JW Jr (1969) A nonlinear mapping for data structure analysis.
\newblock Computers, IEEE Transactions on 100: 401--409.
\bibAnnoteFile{sammon69nonlinear}

\bibitem{r10r}
{R Development Core Team} (2010) {R: A Language and Environment for Statistical
  Computing}.
\newblock R Foundation for Statistical Computing, Vienna, Austria.
\newblock \urlprefix\url{http://www.R-project.org}.
\newblock {ISBN} 3-900051-07-0.
\bibAnnoteFile{r10r}

\end{thebibliography}


\section*{Figure Legends}

\section*{Tables}

\begin{table}[!ht]
\caption{
\bf{Samples and classes (T: tissues, CL: cell lines) for BM1 data set}}
\begin{tabular}{clrl}
\hline
Type & Class & Samples & Notes \\
\hline
T&adipose & 7 & \\
T&brain & 7 & \\
T&breast &  4 & \\
T&colon & 7 &\\
T&heart & 7 &\\
T&liver & 7 &\\
T&lymph node & 7 & \\
T&mixed brain & 4 & \\
T&skeletal muscle & 7 & \\
T&testes & 7 & \\
T&cerebellum & 10 & 6 biological samples: 2 x 1 lane, 4 x 2 lanes\\
\hline
CL&BT474 & 7 & Breast tumor\\ 
CL&HME & 7 & Mammary epithelial\\
CL&MB435 & 4 & Estrogen-ind. breast carcinoma \\
CL&MCF7 & 4 & Breast adenocarcinoma \\
CL&T47D & 4 & Ductal breast epithelial tumor \\
CL&mixed human cell lines & 3 & Breast tumor \\
\hline
\end{tabular}
\begin{flushleft}For each class, a virtual sample is added, consisting of the merged lanes.
\end{flushleft}
\label{tab:data_bm1}
\end{table}

\begin{table}[!ht]
\caption{
\bf{Alignment A1: ranking of the samples of each class in decreasing MCC order}}
\begin{tabular}{ll}
\hline
Class & Rank \\
\hline
mixed brain & 1,2,3,4,\textit{6} \\
mixed human cell lines & 5,7,8,\textit{32} \\
cerebellum & 9,20,33,\textit{35},36,39,\textit{42},\textit{50},51,56,62,\textit{63},67,79 \\
brain & 10,11,12,13,15,16,17,\textit{113} \\
breast & 14,18,19,23,\textit{59} \\
BT474 & 21,22,24,25,26,27,29,\textit{107} \\
MCF7 & 28,31,34,47,\textit{97} \\
lymph node & 30,37,38,40,45,49,54,\textit{115} \\
heart & 41,43,44,46,48,53,60,\textit{117} \\
HME & 52,55,61,64,66,70,76,\textit{118} \\
liver & 57,68,69,71,73,74,77,\textit{116} \\
T47D & 58,65,72,78,\textit{109} \\
testes & 75,80,81,82,84,85,87,\textit{121} \\
colon & 83,86,88,89,90,91,92,\textit{120} \\
adipose & 93,94,95,98,100,103,104,\textit{122} \\
MB435 & 96,99,101,102,\textit{119} \\
skeletal muscle & 105,106,108,110,111,112,114,\textit{123} \\
\hline
\end{tabular}
\begin{flushleft}Italic: merged lanes virtual sample.
\end{flushleft}
\label{tab:pos_bwa}
\end{table}

\begin{table}[!ht]
\caption{
\bf{Alignment A2: ranking of the samples of each class in decreasing MCC order}}
\begin{tabular}{ll}
\hline
Class & Rank \\
\hline
mixed brain & 1,2,3,6,\textit{7} \\
cerebellum & 4,5,\textit{8},9,\textit{12},14,18,19,27,\textit{29},30,31,\textit{32},39 \\
brain & 10,11,13,16,17,20,24,\textit{105} \\
breast & 15,21,23,28,\textit{37} \\
mixed human cell lines & 22,25,26,\textit{115} \\
BT474 & 33,34,35,36,38,40,42,\textit{92} \\
MCF7 & 41,44,46,50,\textit{83} \\
T47D & 43,49,53,54,\textit{89} \\
lymph node & 45,47,48,51,52,58,59,\textit{106} \\
heart & 55,57,60,61,62,64,65,\textit{118} \\
testes & 56,63,66,68,71,74,79,\textit{120} \\
HME & 67,69,72,73,75,77,84,\textit{116} \\
liver & 70,76,78,80,81,82,86,\textit{117} \\
adipose & 85,90,93,98,99,102,107,\textit{122} \\
colon & 87,88,91,94,96,97,103,\textit{121} \\
MB435 & 95,100,101,104,\textit{119} \\
skeletal muscle & 108,109,110,111,112,113,114,\textit{123} \\
\hline
\end{tabular}
\begin{flushleft}Italic: merged lanes virtual sample.
\end{flushleft}
\label{tab:pos_bowtie}
\end{table}

\begin{table}[!ht]
\caption{
\bf{Alignment A3: ranking of the samples of each class in decreasing MCC order}}
\begin{tabular}{ll}
\hline
Class & Rank \\
\hline
heart & 1,2,3,4,5,6,9,\textit{114} \\
brain & 7,8,10,11,12,13,14,\textit{108} \\
liver & 15,16,17,18,19,22,23,\textit{111} \\
skeletal muscle & 20,21,24,25,26,28,32,\textit{123} \\
mixed brain & 27,31,38,40,\textit{104} \\
testes & 29,37,45,49,60,61,69,\textit{119} \\
cerebellum & 30,33,34,35,39,82,85,92,93,103,\textit{106},\textit{107},\textit{109},\textit{110} \\
BT474 & 36,41,43,48,52,56,59,\textit{113} \\
HME & 42,44,46,47,51,54,70,\textit{117} \\
lymph node & 50,53,63,75,79,80,86,\textit{116} \\
colon & 55,57,64,66,67,68,72,\textit{121} \\
adipose & 58,62,65,71,73,77,88,\textit{122} \\
mixed human cell lines & 74,83,87,\textit{105} \\
MB435 & 76,78,81,84,\textit{120} \\
MCF7 & 89,90,91,95,\textit{115} \\
breast & 94,97,99,100,\textit{112} \\
T47D & 96,98,101,102,\textit{118} \\
\hline
\end{tabular}
\begin{flushleft}Italic: merged lanes virtual sample.
\end{flushleft}
\label{tab:pos_tophat}
\end{table}

\begin{table}[!ht]
\caption{
\bf{Alignment A1: elements (classwise) of the 24 clusters emerging by cutting the whole dendrogram at height 2.}}
\begin{tabular}{rl}
\hline
Cluster & Elements \\
\hline
1 &  \textbf{Brain} \\
2 &  \textbf{Heart}, \textbf{Liver} \\
3 & 1 Mixed brain \\
4 & 3 Mixed brain \\
5 & 1 Breast, 3 MCF7, \textbf{T47D}, 5 Testes \\
6 & 1 Cerebellum \\
7 & 3 Cerebellum \\
8 & 5 Lymph node, 1 Testes \\
9 & 5 Colon, \textbf{Adipose} \\
10 & 3 Breast, 1 MCF7, 2 Lymph Node \\
11 & 5 HME, 2 Colon, \textbf{Skeletal muscle} \\
12 & \textbf{Mixed human cell lines} \\
13 &  \textbf{BT474}, 2 HME \\
14 & 1 Testes, \textbf{MB435} \\
15 & \textit{Mixed brain}, \textit{Mixed human cell lines} \\
16 & \textit{Liver}, \textit{Heart} \\
17 & 1 Cerebellum \\
18 & 2 \textit{Cerebellum}, 1 Cerebellum \\
19 & 2 \textit{Cerebellum}, 1 Cerebellum \\
20 & \textit{MB435}, \textit{Testes} \\
21 & 1 Cerebellum, \textit{Breast}, \textit{Brain} \\
22 & 1 Cerebellum, \textit{MCF7}, \textit{BT474}, \textit{T47D} \\
23 & \textit{Skeletal muscle}\\
24 & \textit{Lymph node}, \textit{HME}, \textit{Colon}, \textit{Adipose} \\
\hline
\end{tabular}
\begin{flushleft}Bold: all elements of the class are included in the cluster; Italic: cluster includes the merged lanes virtual sample.
\end{flushleft}
\label{tab:cluster_bwa}
\end{table}

\begin{table}[!ht]
\caption{
\bf{Alignment A2: elements (classwise) of the 23 clusters emerging by cutting the whole dendrogram at height 2.}}
\begin{tabular}{rl}
\hline
Cluster & Elements \\
\hline
1 & 3 Mixed brain \\
2 &  \textbf{Heart}, \textbf{Liver} \\
3 & 1 Mixed brain \\
4 & \textbf{Brain} \\
5 & \textbf{BT474} \\
6 & 5 HME, 1 Adipose, 2 Colon, \textbf{Skeletal muscle}\\
7 & 1 Breast, \textbf{Mixed human cell lines}, 5 Lymph node, 1 Testes \\
8 & 3 Breast, 1 MCF7, \textbf{T47D}, 2 Lymph node, 1 Testes \\
9 & 2 HME, 6 Adipose, 5 Colon \\
10 & 1 Cerebellum \\
11 & 3 Cerebellum \\
12 & 1 Cerebellum \\
13 & \textit{Mixed brain}, 2 \textit{Cerebellum}, 1 Cerebellum \\
14 & 1 Testes, \textbf{MB435} \\
15 & 3 MCF7, 4 Testes \\
16 & \textit{Liver}, \textit{Heart} \\
17 & \textit{Brain}, \textit{Mixed human cell lines} \\
18 & 1 Cerebellum \\
19 & 1 Cerebellum, \textit{Breast}, \textit{T47D}, \textit{BT474} \\
20 & \textit{MB435}, \textit{Testes} \\
21 & 2 \textit{Cerebellum}, 2 Cerebellum \\
22 & \textit{MCF7}, \textit{Lymph node} \\
23 & 1 \textit{HME}, 1 \textit{Colon}, 1 \textit{Adipose}, 1 \textit{Skeletal muscle} \\
\hline
\end{tabular}
\begin{flushleft}Bold: all elements of the class are included in the cluster; Italic: cluster includes the merged lanes virtual sample.
\end{flushleft}
\label{tab:cluster_bowtie}
\end{table}

\begin{table}[!ht]
\caption{
\bf{Alignment A3: elements (classwise) of the 32 clusters emerging by cutting the whole dendrogram at height 2.}}
\begin{tabular}{rl}
\hline
Cluster & Elements \\
\hline
1 & \textbf{Brain} \\
2 & \textbf{Skeletal muscle} \\
3 & 1 Heart \\
4 & 5 Heart, 1 Liver \\
5 & 1 Heart, 6 Liver \\
6 & 1 Mixed brain \\
7 & 3 Mixed brain \\
8 & 5 BT474, 6 HME \\
9 & 1 Testes \\
10 & \textbf{Colon}, 2 BT474, \textbf{Adipose}, 1 HME \\
11 & 4 Breast, \textbf{T47D} \\
12 & 3 MCF7 \\
13 & \textbf{MB435}, 1 MCF7 \\
14 & 1 Testes, 1 Lymph node \\
15 & 6 Lymph node \\
16 & \textbf{Mixed human cell lines} \\
17 & 5 Testes \\
18 & 2 Cerebellum \\
19 & 2 Cerebellum \\
20 & 2 Cerebellum \\
21 & 1 Cerebellum \\
22 & 1 Cerebellum \\
23 & 2 Cerebellum \\
24 & \textit{Testes} \\
25 & 2 \textit{Cerebellum} \\
26 & \textit{Liver}, \textit{Heart} \\
27 & \textit{Mixed human cell lines} \\
28 & \textit{Breast}, \textit{MCF7}, \textit{Lymph node}, \textit{T47D}, \textit{MB435} \\
29 & \textit{BT474}, \textit{HME}, \textit{Colon}, \textit{Adipose} \\
30 & 2 \textit{Cerebellum} \\
31 & \textit{Brain} \\
32 & \textit{Skeletal muscle} \\
\hline
\end{tabular}
\begin{flushleft}Bold: all elements of the class are included in the cluster; Italic: cluster includes the merged lanes virtual sample.
\end{flushleft}
\label{tab:cluster_tophat}
\end{table}

\end{document}